\documentclass[twocolumn,english]{revtex4}
\usepackage{graphicx}

\begin{document}

\title{Late Arriving Particles in Cosmic Ray Air Showers and \\
AGASA's Determination of UHECR Energies}

\author{Hans-Joachim Drescher$^{1,2}$ }

\author{Glennys R. Farrar$^{1}$}

\address{(1) Center for Cosmology and Particle Physics, \\
Department of Physics, New York University, \\
4 Washington Place, New York, NY 10003\\
\\
(2) Frankfurt Institute for Advanced Studies  \\
Johann Wolfgang Goethe-Universit\"at \\
Max-von-Laue Str. 1, 60438 Frankfurt am Main, Germany}

\begin{abstract}

We give the first detailed study of the arrival time distribution of nucleons in UHECR air showers.   We analyze in detail the influence of late arriving particles on the energy determination of the 
AGASA experiment, as well as how the arrival time distribution changes with distance from shower core.   Our calculations are consistent with experimental observations of the AGASA group.  Crucial to obtaining agreement, is the correct implementation of the energy loss for low-energy protons.  We confirm AGASA's estimation of the error in their energy determination associated with late-arriving particles, assuming primary protons.
\end{abstract}
\maketitle

\section{Introduction}

The results of the AGASA collaboration have been followed with great
interest throughout the years because they had the largest exposure
for highest energy cosmic rays. As of now, AGASA reports about 10
cosmic ray events with an energy greater than $10^{20}$
eV\cite{Takeda:2002at}, well beyond the predicted GZK cutoff due to
interaction of particles with the cosmic microwave background,
assuming uniformly distributed sources. By contrast, the HiRes
collaboration reports that their monocular results are consistent with
a GZK cutoff\cite{Abu-Zayyad:2002ta}.  The statistics are not yet
sufficient to decide whether this is a significant discrepancy or not,
but the question of whether there is a systematic shift between the
relative energy normalizations of AGASA and HiRes has assumed special
importance.  AGASA is an extensive air shower array detector while
HiRes uses the air fluorescence technique. 

In this paper we investigate
the contribution of nucleons to AGASA's energy determination, which
has been neglected in previous analyses. The energy determination in
AGASA is based on the particle density, 
measured with plastic scintillator detectors. The reconstructed value
of the signal 
at a distance of 600m from the core is used as an energy estimator,
since this is considered to be rather insensitive to fluctuations and
to not depend too much on the air shower model.  The density of nucleons
at 600m is not negligible. Neutrons (unlike protons) have
a small energy deposit in plastic scintillator but they scatter
elastically with protons in the scintillator.  The
arrival-time-delay distribution of neutrons is very broad and, due to
AGASA's technique for recording the signal, late arriving 
particles have an exponentially increased weight as is explained
below.  This is the motivation for our scrutiny of the issue of late
arriving particles in general and nucleons in particular.  HiRes'
energy measurement entails an entirely different set of issues and
systematic uncertainties which we do not address in this paper. 

\section{Ground nucleons}
\label{gndnucs}
\begin{figure}
\begin{center}\includegraphics[%
  width=1.0\columnwidth]{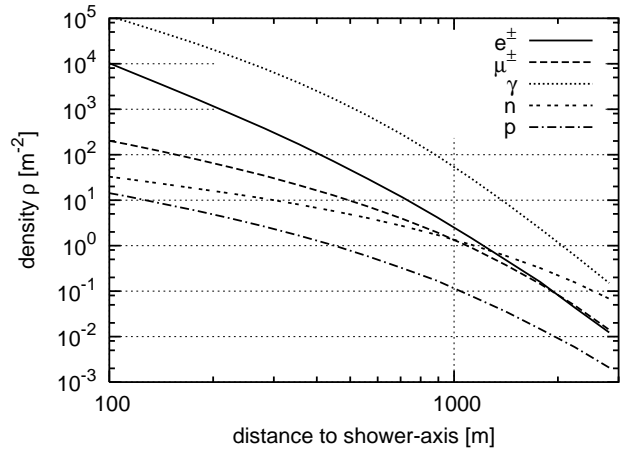}\end{center}

\caption{\label{fig:lat}Lateral distribution of particle density as a function
of distance from the core in a $1\times10^{19}$eV proton-induced
shower. }
\end{figure}
\begin{figure}
\begin{center}\includegraphics[%
  width=1.0\columnwidth]{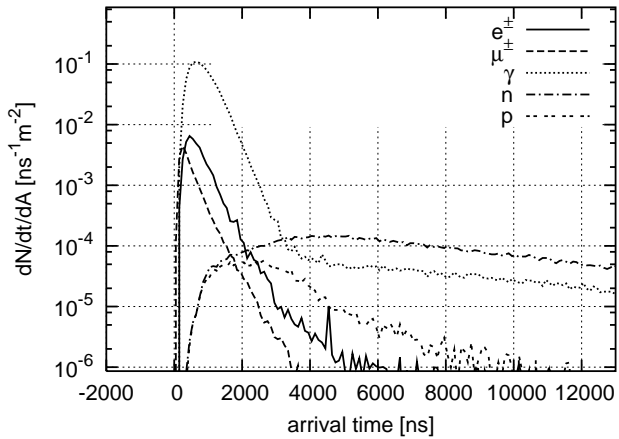}\end{center}

\caption{\label{fig:arrival_time} Arrival time distribution for particles
at 600 m from the shower-core. Times are with respect to
the arrival of the shower-core.}
\end{figure}

The air-showers analyzed in this paper are simulated with the SENECA
model, as introduced in \cite{Drescher:2002cr}. The basic features
of this model are the high energy hadronic model QGSJET01 \cite{QSJET},
the electro-magnetic shower model EGS4 \cite{EGS4}, and different
choices of the low-energy hadronic model: GHEISHA \cite{GHEISHA},
which was the default option in the CORSIKA model, G-FLUKA \cite{FLUKA} and GCALOR
\cite{GCALOR} used in the framework of the GEANT 3.21\cite{GCALOR,GEANT}
package for detector simulation. The model also has some new simulation
techniques which speed up the computation of air showers considerably,
by using the approach of cascade equations as described in detail
in ref. \cite{Bossard:2000jh}. The pure physics content (hadronic
and electromagnetic modelling) using GHEISHA as low-energy hadronic
model, is identical to the one of CORSIKA \cite{CORSIKA}, and the
results are in good agreement with this model.

Figure \ref{fig:lat} shows the averaged lateral distribution of particle
densities as a function of the distance from the shower core, generated
by $1\times10^{19}$ eV vertical proton showers at an altitude 667m above sea
level, using GCALOR as the low-energy hadronic model. All particles
have been plotted which have kinetic energies > 1 MeV. Neutrons become quite
prominent at large distances from the shower-core, unlike protons
which are more readily absorbed in the atmosphere due to ionization
energy loss. 

Electromagnetic particles and muons arrive within about a micro-second,
whereas nucleons can be retarded by many tens of microseconds due to multiple
scatterings. The arrival time distribution of particles of various
species, 600 m from the core, is shown in Fig. \ref{fig:arrival_time}.
Arrival times are measured with respect to the arrival of the shower
core.

\section{Particle Measurement in AGASA}
\label{partmeas}
\begin{figure}
\begin{center}\includegraphics[%
  width=1.0\columnwidth]{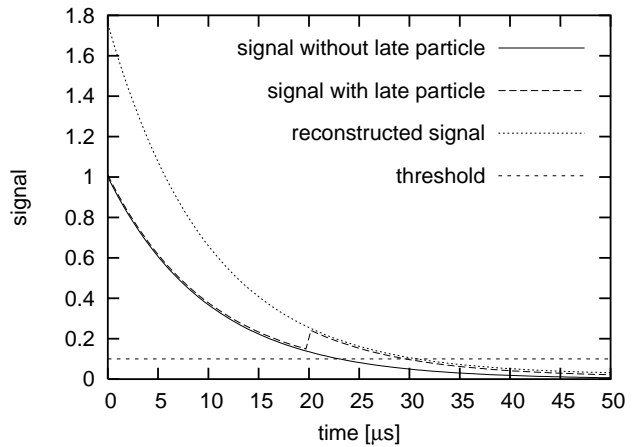}\end{center}

\caption{\label{fig:late}How late arriving particles depositing 10\% of the
initial energy deposit can fake a signal 1.7 times higher.}
\end{figure}

A particle of a specified type, kinetic energy and incidence angle,
gives a certain energy deposit in AGASA's 5 cm thick plastic scintillator.
When a charged particle passes through the dectector, the signal from the photons generated in the scintillator and detected
by the photomultiplier is processed to produce a signal which decays
exponentially, with a time constant of $10\mu s$. The time duration
in which the signal has an amplitude greater than a given discrimination level
is called the pulse-witdh.  
Recording
the pulse-width rather than the signal itself has the advantage that
a large dynamic range of energy deposit is measurable, since the pulse-width
is proportional to the logarithm of the energy deposit. 

An important assumption is that all particles arrive in a time much shorter than
the decay constant of the signal. The bulk of particles arrive in
less than a micro-second, but neutrons can scatter for a much
longer time since they do not lose energy by ionization of air molecules,
as is reflected in Fig. \ref{fig:arrival_time}. The effect of a particle
arriving 20$\mu\mathrm{s}$ late, with an energy deposit 10\% as large
as the original signal, is shown in Fig. \ref{fig:late}. Even though
the late energy deposit in this particular example is small, the time delay
increases its importance by a factor of $\exp(20\mu\mathrm{s}/10\mu\mathrm{s})=7.4$
and the inferred total energy deposit is overestimated by roughly a factor
of 1.7 .

One can easily calculate the effect of this time delay for the measurement
of particle densities. It increases the signal by the factor 
\begin{equation} \label{f}
f=\frac{\int_{t_{0}}^{t_{1}}\frac{dE_{\mathrm{dep}}}{dt}\exp((t-t_{0})/10\mu s)dt}{\int_{t_{0}}^{t_{2}}\frac{dE_{\mathrm{dep}}}{dt}dt},
\end{equation}
where $t_{0}$ is the time when the detector triggered and $t_{1}$
is the time when the signal falls below the threshold or 128$\mu\mathrm{s}$,
whichever is less. $t_{2}$ represents the arrival time of the last
particle, not necessarily equal to $t_{1}$. Thus $f$ is the factor by
which the measured signal has to be reduced in order to obtain the
true energy deposit in a detector. 

\begin{figure}
\begin{center}\includegraphics[%
  width=1.0\columnwidth]{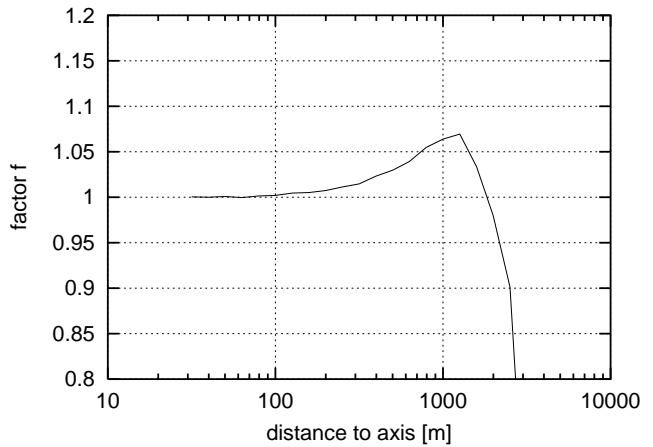}\end{center}

\caption{\label{fig:rad_comp}The effect of time delay on the measurement of
the total energy deposit of electrons, muons and photons.}
\end{figure}

Figure \ref{fig:rad_comp} shows $f$
as a function of radius for the total signal of electrons, muons and
photons, using the shower simulation described in section \ref{gndnucs}. The factor $f$ increases as a function of the radius due
to greater spreading of the arrival times at large distance. At very large distances
$f$ decreases again, when the particle number in a detector falls to of order unity. If a single particle arrives at a detector, $f=1$ as is evident from replacing $dE_{\mathrm{dep}}/dt$ in eq. (\ref{f}) with a delta function at $t_0$.  However
$f$ can also be less than 1, since in general the arrival time
$t_{2}$ can be larger than the time $t_{1}$ when the pulse falls below
the threshold. If for example a second particles arrives after $t_{1}$
the AGASA acquisition system generates a second signal but it is 
ignored in the analysis.  Effects arising when the particle
number is of order one are a function of the area of the detector
and primary energy as well as distance to the core. The AGASA collaboration measured the effect of the arrival time distribution, in a 30 m${}^{2}$ detector with time-resolution as described in ref.
\cite{Takeda:2002at}.

\section{Effect of Neutrons and Protons in a Plastic Scintillator}

The energy-deposit of neutrons in a 5cm-plastic scintillator is calculated
with the GEANT-package \cite{GEANT} and the Gcalor interface \cite{GCALOR}
relevant for low energy neutrons below 10 GeV. Above 10 GeV, routines
from the FLUKA package are called automatically \cite{FLUKA}. The
geometrical setup consists of a 1.5m~x~1.5m~x~5cm plastic scintillator
in a 1.6mm thick iron box which is placed in a farmer's storage house
made with 0.4mm thick iron plates. We have checked that we can reproduce
the energy deposit spectra used by the AGASA-collaboration for electrons,
positrons, muons and photons \cite{Sak2001}. Protons deposit energy
in a plastic scintillator since they are charged particles. The main
process is ionization. Neutrons affect the scintillator by scattering
elastically with the protons of hydrogen abundantly available in plastic
scintillators;  the recoil protons then deposit energy \cite{birks2}.

\begin{figure}
\begin{center}\includegraphics[%
  width=1.0\columnwidth]{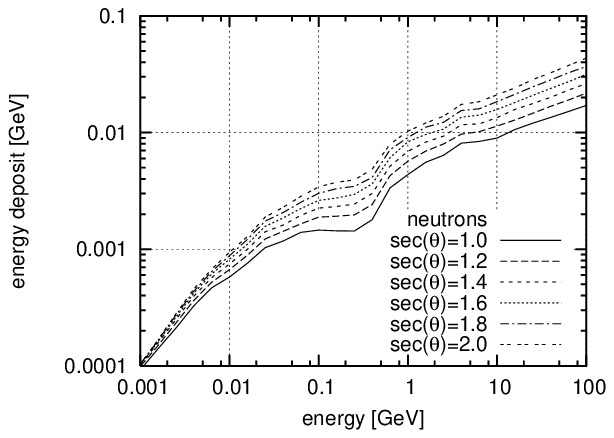}\end{center}

\begin{center}\includegraphics[%
  width=1.0\columnwidth]{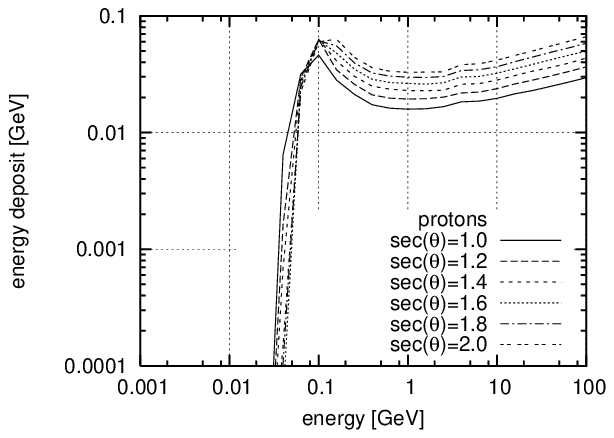}\end{center}

\caption{\label{fig:neu_engy}Energy deposit in a 5cm plastic scintillator
of vertically arriving neutrons (top) and protons (bottom) of different
energies.}
\end{figure}

The energy deposit spectra of neutrons and protons are shown in figure
\ref{fig:neu_engy} for particles arriving at different angles. One
sees the increasing energy deposit due to the longer path length in
the material scaling with $1/\cos(\theta)$. 

\begin{figure}
\begin{center}\includegraphics[%
  width=1.0\columnwidth]{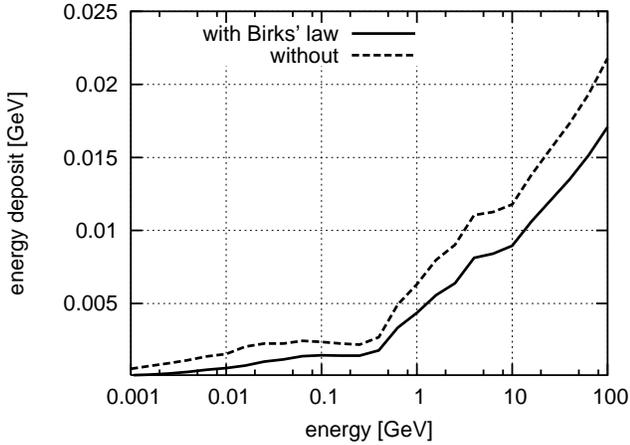}
\end{center}
\caption{\label{fig:birk}The effect of Birks' law on the effective energy
deposition spectrum of neutrons.}
\end{figure}

If the energy deposit in a scintillator is locally concentrated, a
saturation of the scintillation response is observed. The impact of
this on the scintillation signal generated by neutrons and protons
is described by Birks' law \cite{birks1}, and is shown in Fig. \ref{fig:birk}.
Throughout our calculations, the effect of Birks' law has been included,
using the parameters proposed in the GEANT manual which are consistent
with values proposed in \cite{birks1}. The result is expressed by
quoting an effective energy deposit, which would give the equivalent
scintillator response in the absence of this effect.


\section{Calibration of the theory}
\label{sec:Calibration}
\begin{figure}
\begin{center}\includegraphics[%
  width=1.0\columnwidth]{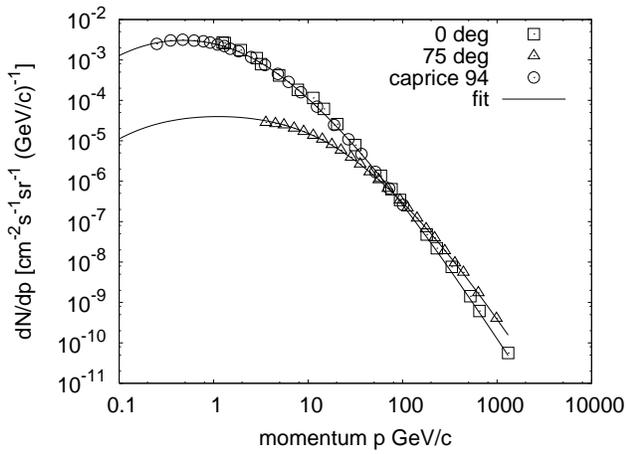}
\end{center}
\caption{\label{fig:atm_muons} The fitted distributions of atmospheric muons.}
\end{figure}
\begin{figure}
\begin{center}\includegraphics[%
  width=1.0\columnwidth]{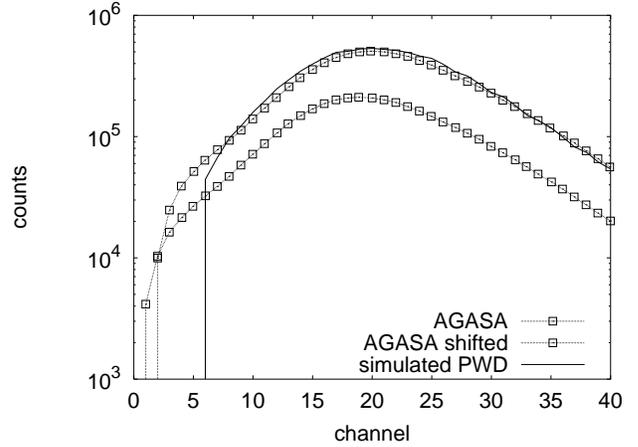}
\end{center}

\caption{\label{fig:PWD} Theory calibration according to the experiment. Each
channel corresponds to $500\textrm{ns}$. AGASA calibrates
with the peak value $t_{1}$ whereas we calibrate the threshold 
of the signal (energy deposit) such that the peak is at
$t_{1}=10\mu\textrm{s}$.}  
\end{figure}

AGASA calibrates their detectors with the ambient signal from
atmospheric muons, also called "omnidirectional" muons, which are the
dominant source of signals in individual detectors.  Because these muons are
close to being minimum-ionizing, the energy deposited by a single muon
depends primarily on its angle (tracklength) and depends only weakly
on its energy.  Therefore the pulsewidth distribution of an individual
detector is well understood and suitable for calibrating the
scintillator response.  We calibrate the theory in the same way.  

>From data of  refs. \cite{Allkofer:1971qr,Jokisch:1979gh}
and Caprice 94 data \cite{Kremer:1999sg}, a functional fit of the muon
spectrum  
$dD(p,\theta)/dp$ is obtained; data and fit are shown in
Fig. \ref{fig:atm_muons}. Then a Monte Carlo simulation is performed,
in which muons are produced with this distribution and propagated
through 
our detector simulation. Here, it is crucial to apply a smearing function
to the energy deposit to reproduce the experimental shape of the pulse
width distribution.  We have found that a suitable smearing function
is $P(x)\sim\exp(-0.5\ln^{2}(xb)/\sigma^{2})/x$, which can be obtained
by taking the exponential of a random number sampled from a Gaussian
distribution around zero and with width $\sigma$. Empirically, this
function imitates Landau fluctuations in the photo-multiplier tube.
The parameter $b$ is adjusted such that the mean is equal to one:
$\int xP(x)dx/\int P(x)dx=1$.  The values $\sigma=0.3$ and $b=1.045$
proved to describe best the 
shape of the PWD observed by AGASA.  The quality of this fit is shown
in Fig. \ref{fig:PWD}. 

The next step in calibration is to set the absolute normalization of
the signal.   To convert AGASA's calibration to an energy scale we
need to return to a more detailed discussion of their data recording
procedure, which was described schematically in section
\ref{partmeas}. If the signal threshold is set at $V_{\rm thr}$, then 
the time-above-threshold or pulse-width, $t_{\rm P}$, is implicitly given
by \cite{Takeda:2002at} 
\begin{equation}
V_{\rm thr}=V\exp(-t_{\rm P}/\tau)\,\,,  \label{for:Vd}
\end{equation}
where $V$ is the pulse-height of the signal, proportional to the 
scintillation response and therefore to energy deposit. 
A priori, $V$ and $V_{\rm thr}$ are measured in an arbitrary unit relating to
the gain of the PMTs.

By defining $t_{1}$ as the pulse-width of a single particle 
with pulse-height $V_1$ one obtains
\begin{equation}
 V_{\rm thr}=V_1\exp(-t_1/\tau) ~~, \label{for:V1}
\end{equation}
and
\begin{equation}
N=\frac{V}{V_1}=\exp((t_{\rm P}-t_{1})/\tau)  \label{for:N}
\end{equation}
 as the number of particles for a arbitrary pulse-width $t_{\rm P}$.
One can consider $N$ as some effective number of particles, since it 
comprises different particle types. 
AGASA sets their gain and threshold so that the peak of the
pulse-width distribution of atmospheric muons occurs roughly in
channel 20, as shown in Fig.  \ref{fig:PWD}.  This corresponds to
$t_1\approx\tau=10\mu s$ because one channel is 500ns. The precise
calibration is then given by the peak value $t_1$.

In our case the signal is just the energy deposit $E$ in the 
plastic scintillator, so $V \rightarrow E$ applies to all formulas
above.  We calibrate the threshold $E_{\rm thr}$ such that 
the peak of the pulse-width distribution of omnidirectional muons is
exactly at $t_1=\tau = 10\mu$s.  
According to (\ref{for:V1}), the energy deposit of an effective
particle is then $E_1=E_{\rm thr}/\exp(-1)$ and for arbitrary 
$t_{\rm P}$ the effective number of particles is just

\begin{equation}
 ~~ N=\exp(t_{\rm P}/\tau-1)~.
\end{equation}

Our simulation shows that the peak in the energy deposition of the
omnidirectional muons corresponds to $E_1=0.011$ GeV per particle.

\section{Lateral distribution functions}

\begin{figure}
\includegraphics[width=1.0\columnwidth]{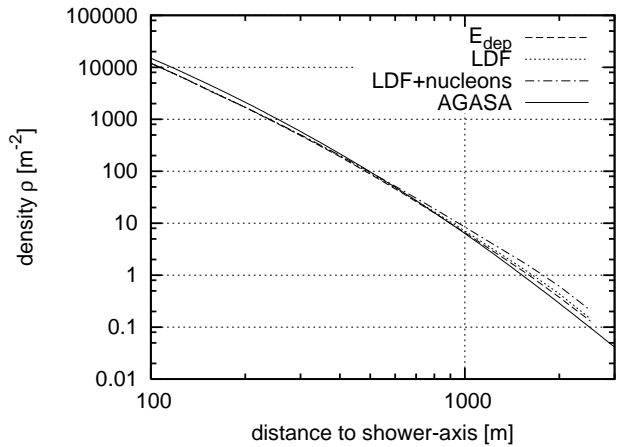}
\caption{\label{cap:LDF_Gheisha} LDF for a $10^{19}$ eV vertical proton induced shower using QGSJET01/GHEISHA.}
\end{figure}

\begin{figure}[h]
\includegraphics[%
  width=1.0\columnwidth]{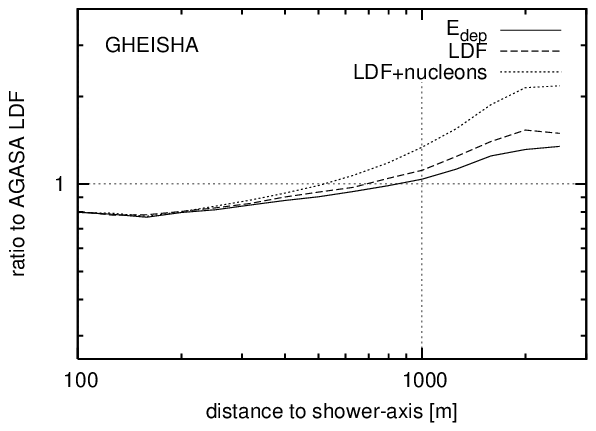}

\includegraphics[%
  width=1.0\columnwidth]{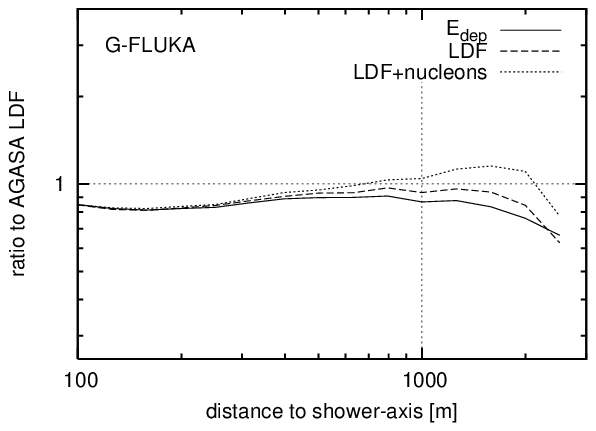}

\includegraphics[%
  width=1.0\columnwidth]{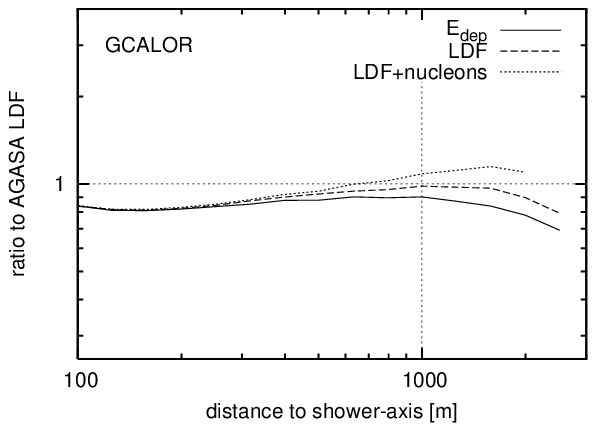}

\caption{\label{cap:LDF_ratio}LDF scaled by the AGASA empirical function
for different low energy models and with the effects of the shower
front and late nucleons.}
\end{figure}

The AGASA collaboration finds experimentally that the lateral distribution of
particle density (as defined via equation (\ref{for:N})) can be described by the
empirical function:  
\begin{equation}
S(R)=C\left(\frac{R}{R_{M}}\right)^{-\alpha}\left(1+\frac{R}{R_{M}}\right)^{-(\eta-\alpha)}\left(1+\left(\frac{R}{1km}\right)^{2}\right)^{-\delta}, \label{for:LDF}
\end{equation}
where the parameters $\alpha=1.2$ and $\delta=0.6$ are fixed,  $R_{M}=91.6m$
is the Moliere radius two radiation lengths above the altitude
of the Akeno observatory, and the slope $\eta$ at distances large compared to $R_{M}$
is an experimentally determined function of the shower zenith angle
$\theta$:  $\eta=3.84-2.15(\sec\theta-1)$ \cite{Takeda:2002at} .

By fitting ``data" from their Monte Carlo shower and detector simulation to the lateral dependence of equation (\ref{for:LDF}), AGASA obtained the conversion formula (formula~(1) from Ref.\cite{Takeda:2002at} )
\begin{equation}
E=2.03\times10^{17}S(600m)\,\mathrm{eV}\, \label{for:AGASA engy},
\end{equation}
using calculations for the Akeno  observatory at the altitude $900$~m. It was realized later that the average altitude
of the whole AGASA array is actually $667$~m, significantly different from that of the Akeno sector only. However when the AGASA
analysis is corrected for the change in average altitude and still other effects (mainly shower front
thickness and delayed particles -- see table~2 from
\cite{Takeda:2002at}), AGASA finds a final conversion formula which is coincidentally
the same as their original one, (\ref{for:AGASA engy}).  Below, when we refer to the AGASA LDF, we mean equation (6) normalized by (\ref{for:AGASA engy}).

We now compare the results of our calculations for various cases, to AGASA's empirical results.  We use an average altitude $667$~m
and the procedure of Sec. \ref{sec:Calibration} to obtain the signal as a function of energy deposit.  Fig. \ref{cap:LDF_Gheisha} shows our predicted LDF for an average
$10^{19}$ eV proton-induced vertical shower compared to the the AGASA
empirical function (\ref{for:LDF}).  To quantify the importance of
including the shower-front thickness and the contribution of nucleons,
the figure shows three cases, all using QGSJET01/GHEISHA as the
hadronic models.  The curves are labeled $E_{dep}$, LDF and
LDF+nucleons, and are computed as follows:
\begin{itemize}
\item $E_{dep}$: Only the energy deposit of electrons/positrons, muons, and photons is included; particles are assumed to arrive simultaneously.
\item LDF:  Same particles contributing as above, but with non-trivial distribution of shower front thickness taken into account.
\item LDF+nucleons: the above, and additionally the effect of 
                    nucleons and their time delay taken into account.
\end{itemize}

One observes that the combination of hadronic models 
QGSJET01/GHEISHA, gives a slightly too flat slope of the LDF in all
cases. This has already been found in ref. \cite{Nagano:1999xk} based
on ``$E_{dep}$" alone. We see here that including the shower front thickness
and the nucleons increases the discrepancy. This can be 
seen better in Fig. \ref{cap:LDF_ratio} where the results are shown as a ratio
to the AGASA LDF.  The lower panels of this figure show the LDF as obtained
by choosing G-FLUKA and GCALOR as low energy hadronic model instead of
GHEISHA. One sees that these two give a better description of the
AGASA LDF, though still dropping slightly more slowly with distance. That
the tails of LDFs can be sensitive to the low energy hadronic model has already been
shown in ref. \cite{Drescher:2002vp,DrescherTsukuba,Drescher:2003gh}. 

The values of the scaled LDFs at $600$ m in Fig. \ref{cap:LDF_ratio} 
allow us to infer the deviation of AGASA's energy estimation equation (\ref{for:AGASA engy}) from that implied by our simulations. Table \ref{tab:ratio} shows the ratio of our predicted signal to that ``predicted" by AGASA using (\ref{for:AGASA engy}) as conversion factor, with QGSJET01 for the high energy model in each case.  Thus for a given observed signal, if GHEISHA were the correct low-energy hadronic model AGASA would overestimate the energy of a vertical proton by about 5\%, while if G-FLUKA or GCALOR correctly describe the low energy interactions, the AGASA energy estimate would be about 2\% too low, when all the effects of late-arrivers and shower-front thickness are included.

\begin{table}
\begin{tabular}{|c|c|}
\hline
Model&
deviation from (\ref{for:AGASA engy}) \tabularnewline
\hline
QGSJET01/GHEISHA&1.049\tabularnewline
\hline
QGSJET01/G-FLUKA&0.983\tabularnewline
\hline
QGSJET01/GCALOR&0.978\tabularnewline
\hline
\end{tabular}
\caption{\label{cap:table}\label{tab:ratio}
The deviation of the energy estimation obtained in these simulations.}
\end{table}

\section{Comparison of Predicted and Observed Arrival Time Distributions}

\begin{figure}
\includegraphics[width=1.0\columnwidth]{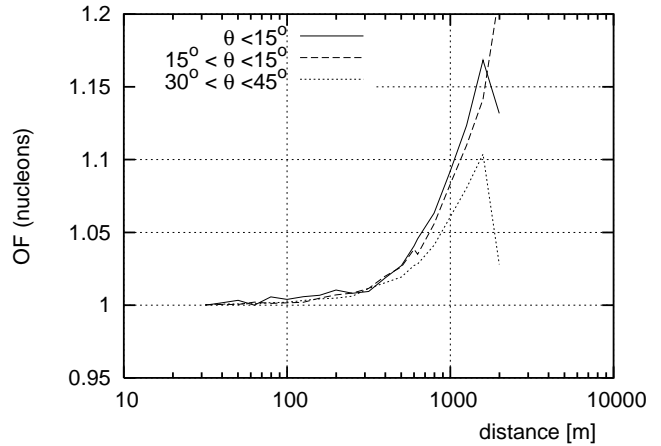}
\caption{\label{cap:OF}Over-estimation factor due to late arriving nucleons as a function of radius, for $10^{19}\textrm{eV}$ proton-induced showers.}
\end{figure}

As discussed in Sec. \ref{gndnucs}, given AGASA's method of recording the signal, late arriving particles such as nucleons could in principle distort the determination of the primary energy; Fig. \ref{cap:OF} shows the simulated value of the overestimation factor $f$ due to late arriving nucleons as a function of distance to the core.  However as we saw in the previous section, the impact of late arrivers is compensated by other effects which had originally been neglected in AGASA's calculations, so that we find the average AGASA energy determination to be quite good, assuming the primaries are protons and QGSJET01 is an adequate high energy model.  The overestimation factor at 600m is not more than 5\% for a vertical $10^{19}\textrm{eV}$ protons and we have confirmed it is the same for $5\times10^{19}\textrm{eV}$ primaries.

The arrival time distribution itself has not up to now been critically compared to observations, which we do in this section.  In addition to the standard detectors, the AGASA collaboration\cite{Takeda:2002at} used a 30 $m^2$ detector in combination with the rest of the AGASA array to record the actual signal in that detector as a function of time, within $\pm ~30 \mu$s of the first particles of the shower.  Ref. \cite{Takeda:2002at} remarked that including nucleons
in model calculations gives an LDF which is much too flat compared to
what is observed  experimentally.  We have found that this behavior can be produced by an imprecise implementation of energy loss.

\begin{figure}
\includegraphics[width=1.0\columnwidth]{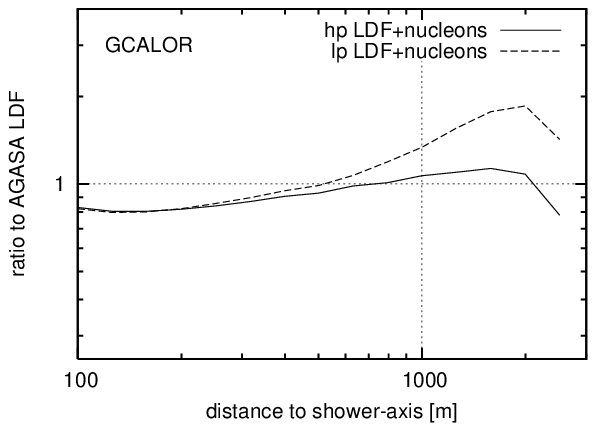}
\includegraphics[width=1.0\columnwidth]{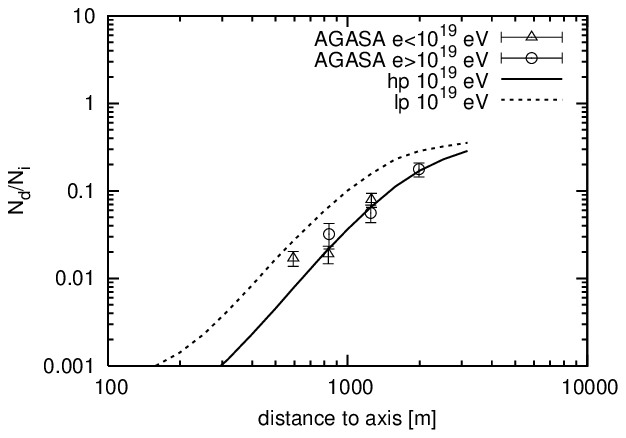}
\caption{\label{cap:ldf_arrival}LDF (upper panel) and fraction 
of delayed particles (lower panel) with high precision (hp) 
and low precision (lp) calculations of energy loss of charged particles.}
\end{figure}

The Bethe Bloch 
formula for the energy loss of charged particles diverges for values
of the relativistic gamma-factor approaching $\gamma=1$. When
particles are propagated  over a large distance in a single step, this
divergence in $dE/dx$ is not properly taken into account if the
simulation calculates $dE/dx$ at the starting point only and takes it
to be the same over the total step length. If one takes 
for the step length the distance to the next
interaction or decay, it is for non-relativistic particles often
much too long for constant $dE/dx$ to be a good approximation.  For
this reason the energy loss of protons, and neutrons coming from them
as secondaries in the shower, is
underestimated.  A simple cure for this problem is to limit the step
length to a suitably low value $l_{\max}$. We found 
that for $l_{\max}\leq10$~m the results are stable, i.e., the low energy
spectra of hadrons do not change any more as $l_{\max}$ is reduced further. 

The influence of the higher precision achieved can be seen 
in the upper panel of Fig. \ref{cap:ldf_arrival} where
the LDF, including nucleons, has been calculated with high- and low-precision
energy loss.  ``Low-precision" means taking the step length
as proposed by the total cross-section of the interaction, usually
several hundred meters at ground level, as done in standard 
simulations.  It can be seen that the low precision energy loss calculation 
results in a significant overestimation of the LDF at large distances. 

The change in precision is not important for particles other than nucleons.  
Energy loss is accurately treated in electromagnetic shower codes, and 
when other charged particles such as muons and pions in the shower 
become non-relativistic they decay too quickly for their subsequent 
energy loss to matter.  All results shown in this paper have been 
computed with high precision, for all charged particles in the shower.

Now we compare the observed delay time to our simulations.  The lower panel of Fig. \ref{cap:ldf_arrival} shows the
fraction of delayed particles as a function of the distance from the
shower-axis, calculated with high and low precision. The data is from the
AGASA collaboration measurement discussed above\cite{Takeda:2002at} 
and includes isotropically distributed showers up to $45^\circ$ incident angle.
The simulations are done in the same way, assuming purely proton primaries.
The fraction of delayed particles is defined as the energy deposit
of particles arriving later than $3~ \mu$s after the first particles
in the detector, divided by the energy deposit of all particles. 
The late arrivers in the high precision calculation are 
reduced by a factor of two, 
and the results are in good agreement with the data.

\section{Summary and Conclusions}
We have analyzed the effect of late arriving particles on the 
energy determination of the AGASA experiment.  We find that the net effect of the arrival time spread in the electromagnetic and muonic components of the shower, and the inclusion
of nucleons in the simulation, has only a small effect on the AGASA
energy determination.  It shifts the true energy compared to the
energy as determined without these effects, upward by a few percent in
the models which give the best agreement with the shape of the lateral distribution function.  This agrees with the correction applied by AGASA\cite{Takeda:2002at}.  The LDF measured by AGASA is best described with G-FLUKA
or GCALOR as the low energy hadronic model; the corresponding LDF using
the GHEISHA code is too flat. 

We have demonstrated for the first time that shower simulations can describe the distribution of arrival times of late-arriving particles, primarily nucleons, as observed by the AGASA collaboration\cite{Takeda:2002at}.  When analyzing neutrons and protons in air shower simulations, the energy loss of charged particles has to be done with care. It is straightforward but crucial to compute these energy losses accurately, otherwise the contribution of these particles to the signal can be greatly overestimated.  The Pierre Auger Observatory records detailed arrival time information in each tank, so accurately modeling the observed arrival time distributions should be a powerful new tool for validating shower simulations.  An interesting question, left to the future, is to what extent the arrival time distribution can be useful for studying the composition of UHE cosmic rays.

\subsection*{Acknowledgments}

HJD acknowledges support from NASA grant NAG 9246, and from the
German BMBF/DESY grant 05CT2RFA/7; GRF is supported by NASA grant NAG-9246 
and NSF-PHY-0101738. 

Computations were done on computer clusters at NYU financed in part
by the Major Research Instrumentation grant NSF-PHY-0116590 and 
at the Center for Scientific Computing, Frankfurt am Main. 

This research would not have been possible without M. Teshima's
willingness to explain details of the AGASA detector and its signal
processing procedures. We have also benefited from 
discussions with and assistance from many other colleagues including
R. Djilbikaev, J. Sculli, A. Watson, S. Westerhoff, and Rex Tayloe.

\end{document}